\documentstyle[12pt]{article}
\begin{document}\bibliographystyle{plain}\begin{titlepage}
\renewcommand{\thefootnote}{\fnsymbol{footnote}}\hfill\begin{tabular}{l}
HEPHY-PUB 753/02\\UWThPh-2002-8\\June 2002\end{tabular}\\[4cm]\Large
\begin{center}{\bf ELECTRIC POLARIZABILITY OF MESONS\\IN SEMIRELATIVISTIC
QUARK MODELS}\\\vspace{2cm}\large{\bf Wolfgang LUCHA\footnote[1]{\normalsize\
{\em E-mail address\/}: wolfgang.lucha@oeaw.ac.at}}\\[.3cm]\normalsize
Institut f\"ur Hochenergiephysik,\\\"Osterreichische Akademie der
Wissenschaften,\\Nikolsdorfergasse 18, A-1050 Wien, Austria\\[1cm]\large{\bf
Franz F.~SCH\"OBERL\footnote[2]{\normalsize\ {\em E-mail address\/}:
franz.schoeberl@univie.ac.at}}\\[.3cm]\normalsize Institut f\"ur Theoretische
Physik, Universit\"at Wien,\\Boltzmanngasse 5, A-1090 Wien, Austria\vfill
{\normalsize\bf Abstract}\end{center}\normalsize The electric polarizability
of mesons, in particular, that of the charged pion, is studied in the
framework of a semirelativistic description of hadrons as bound states of
valence quarks in terms of a Hamiltonian composed of the relativistic kinetic
energy as well~as~a phenomenological potential describing the strong
interactions between the quarks.~The quark-core contribution to the electric
polarizability obtainable in quark models of~this kind is in the
semirelativistic approaches even smaller than in the nonrelativistic
limit.\vspace{3ex}

\noindent{\em PACS numbers\/}: 12.39.Ki, 12.39.Pn, 13.40.Em
\renewcommand{\thefootnote}{\arabic{footnote}}\end{titlepage}

\normalsize

\section{Introduction}An elementary particle may be considered as
fundamental, and thus called ``pointlike,'' if it does not exhibit an
experimentally observable internal structure, i.e., if it does~not have a
discernible spatial extension and is not composed of detectable
subcomponents. Nonvanishing electromagnetic polarizabilities of a particle,
on the other hand, provide a clear empirical evidence that this particle
cannot be fundamental in the above sense. Rather, any particle with this
property has to be regarded as a {\em composite\/} system.~More precisely, it
must be built up of subcomponents carrying nonvanishing electric charges.

The electromagnetic polarizabilities of a particle characterize the~dipole
moments induced by the presence of an external electromagnetic field. They
therefore constitute fundamental quantities which represent a measure of the
rigidity, stiffness or resistance to deformation of the internal structure of
this composite system upon imposition~of~an external electromagnetic field.
Phrased the other way round, they probe the ease with which a composite
system can be polarized by the external field. Clearly, more detailed
analyses of electromagnetic polarizabilities will then yield still deeper
insights into the internal structure of such a composite system. For
instance, they will allow to estimate the strength of the basic interactions
responsible for the formation of this bound state.

Within the realm of strong interactions, electromagnetic polarizabilities of
hadrons play, for the following reasons, an important r\^ole for any
effective theory proposed as~a realization of quantum chromodynamics (QCD) at
low energies. Chiral symmetry---as implemented, for instance, in chiral
perturbation theory ($\chi$PT)---allows to formulate~a precise, unambiguous
prediction for the electric polarizability of the charged pion. This
prediction establishes a relation between, on the one hand, the electric
polarizability~of the charged pion and, on the other hand, the ratio of
vector and axial-vector structure constants entering in radiative
charged-pion beta~decay by the governing weak current. Consequently, the
electric polarizability of the charged pion provides a stringent test~of the
chiral symmetry and hence of (the low-energy limit of) quantum
chromodynamics.

Usually, the electric polarizability of a particle $P$ is denoted either by
$\alpha_P,$ or by~$\alpha_{\rm E},$ or by $\bar\alpha$ when one discriminates
between the contributions of the classical polarizability (related to the
electromagnetic particle size) and the intrinsic polarizability. However, in
order to avoid confusions with the electromagnetic fine structure constant
$\alpha$ and the strong fine structure constant $\alpha_{\rm s},$ we denote
the electric polarizability by the symbol~$\kappa.$

\subsection{Electric polarizability of pions: theory versus experiment}
\label{Subsect:TVE}In spite of the principal importance of the
electromagnetic polarizabilities just recalled, experimental data for
electric and magnetic polarizabilities of mesons are very rare~and the sparse
results existing now are still not unambiguous. In particular, concerning~the
electric polarizability of the charged pion, the experimental situation is
far from being consistent; consequently, the present state of the art cannot
be regarded as satisfactory. In quantum physics, the appropriate means for
the investigation of the electromagnetic polarizabilities of a given
composite particle $P$ is the analysis of its Compton
scattering$$\gamma+P\rightarrow\gamma+P$$or of related processes obtained by
crossing symmetry, such as the production of a~$P\bar P$ pair in
photon--photon scattering,$$\gamma+\gamma\rightarrow P+\bar P\ ,$$or the
decay of a $P\bar P$ bound state (by the annihilation of $P$ and $\bar P$)
into two photons, $$P+\bar P\rightarrow\gamma+\gamma\ .$$ Our object of
desire is the charged pion, i.e., $P=\pi^\pm.$ The problem, and
simultaneously the experimental challenge, is the fact that a pion target is
not available. Consequently, the particular process under investigation must
be embedded into a more suitable,~that is, experimentally accessible,
reaction. This circumstance renders all the corresponding measurements very
difficult. Compton scattering off charged pions, that is, the
process$$\gamma+\pi^\pm\to\gamma+\pi^\pm\ ,$$may be experimentally realized
either, in form of the Primakoff effect, by {\it radiative~pion scattering\/}
on a nucleus of atomic number $Z$ (involving, of course, a
virtual~photon),$$\pi^-+Z\to\pi^-+Z+\gamma\ ,$$as has been done by an
experiment using the Sigma spectrometer at Serpukhov~\cite{Serpukhov83},~or
by {\it radiative pion photoproduction\/} in photon--nucleon scattering
(involving, of course,~a virtual pion),$$\gamma+{\rm p}\to\gamma+{\rm
n}+\pi^+\ ,$$as has been done by an experiment performed at the electron
synchrotron PACHRA of the Lebedev Physical Institute~\cite{Lebedev86}.
Two-photon production of charged pion pairs,~that is, the
process$$\gamma+\gamma\to\pi^++\pi^-\ ,$$may be experimentally realized in
the {\it electron--positron collision\/} (involving two virtual photons)
$${\rm e}^++{\rm e}^-\to{\rm e}^++{\rm e}^-+\pi^++\pi^-\ ,$$as has been done
by an experiment with the Mark II detector at the SLAC storage~ring
PEP~\cite{MarkII90}. Table~\ref{Tab:Expts} summarizes the results for the
charged-pion electric polarizability~$\kappa_{\pi^\pm}$ reported by the above
experiments. Our weighted average of these measurements~reads\begin{equation}
\kappa_{\pi^\pm}^{\rm exp}=(4.3\pm1.2)\times10^{-4}\;\mbox{fm}^3\
.\label{Eq:exp-av}\end{equation}

\begin{table}[hbt]\caption{Experimental values of the electric polarizability
$\kappa_{\pi^\pm}$ of the charged pions~$\pi^\pm.$ These values have been
extracted from experiments using two different basic processes: the
Serpukhov~\cite{Serpukhov83} and the Lebedev~\cite{Lebedev86} experiments
used Compton scattering off pions whereas the Mark II~\cite{MarkII90} result
has been deduced from the photon--photon cross-section.}\label{Tab:Expts}
\begin{center}\begin{tabular}{lll}\hline\hline\\[-1.5ex]
\multicolumn{1}{c}{Experiment}&\multicolumn{1}{c}{Reaction}&
\multicolumn{1}{c}{$\kappa_{\pi^\pm}\ [10^{-4}\;\mbox{fm}^3]$}
\\[1ex]\hline\\[-1.5ex]Serpukhov 1983
\cite{Serpukhov83}&$\pi^-+Z\to\pi^-+Z+\gamma$&
$6.8\pm1.4\;{\rm(stat.)}\pm1.2\;{\rm(syst.)}$\\[1ex]Lebedev 1986
\cite{Lebedev86}&$\gamma+{\rm p}\to\gamma+{\rm n}+\pi^+$&
$20\pm12\;{\rm(stat.)}$\\[1ex]Mark II 1990 \cite{MarkII90}&${\rm e}^++{\rm
e}^-\to{\rm e}^++{\rm e}^-+\pi^++\pi^-$& $2.2\pm1.6\;{\rm(stat.+syst.)}$
\\[1ex]\hline\hline\end{tabular}\end{center}\end{table}

\noindent
On the theoretical side, electromagnetic polarizabilities of mesons, in
particular, those of the charged pion, have been investigated by various
approaches, namely, within the framework of a naive static nonrelativistic
potential model for the interquark forces~\cite{Leeb}, of a (generalized)
Nambu--Jona-Lasinio model \cite{Bernard89}, of chiral perturbation theory
\cite{ChPT},~of~a so-called ``Dubna quark confinement model''
\cite{Ivanov92}, as well as of a relativistic quark model \cite{Koll01} based
on the instantaneous but quantum-field-theoretic Bethe--Salpeter formalism.
Table~\ref{Tab:Theory} attempts to collect the theoretical predictions for
the electric polarizability~$\kappa_{\pi^\pm}$ of the charged pion obtained
within the various treatments of the pion just mentioned.

\begin{table}[ht]\caption{Theoretical results for the electric polarizability
$\kappa_{\pi^\pm}$ of the charged pions~$\pi^\pm$}\label{Tab:Theory}
\begin{center}\begin{tabular}{ll}\hline\hline\\[-1.5ex]
\multicolumn{1}{c}{Quark bound-state approach}&\multicolumn{1}{c}{Predicted
(range of) $\kappa_{\pi^\pm}\ [10^{-4}\;\mbox{fm}^3]$}\\[1ex]\hline\\[-1.5ex]
nonrelativistic quark potential model~\cite{Leeb}&$0.054$\\[1ex](generalized)
Nambu--Jona-Lasinio model~\cite{Bernard89}&$10.5\div 12.5$\\[1ex]chiral
perturbation theory (at two loops)~\cite{ChPT}&$2.4\pm0.5$\\[1ex]``Dubna
quark confinement model''~\cite{Ivanov92}&$3.63$\\[1ex]relativistic
Bethe--Salpeter quark model~\cite{Koll01}&$4.51\div
6.93$\\[1ex]\hline\hline\end{tabular}\end{center}\end{table}

A brief inspection of the theoretical predictions for the electric
polarizability $\kappa_{\pi^\pm}$~of the charged pion given in
Table~\ref{Tab:Theory} allows to draw immediately the following
conclusions:\begin{itemize}\item The most recent experimental determination
of $\kappa_{\pi^\pm}$ by the Mark II collaboration \cite{MarkII90} reports a
result for $\kappa_{\pi^\pm}$ significantly smaller than the previous
measurements. The inclusion of this value results in a considerable reduction
of the discrepancy between the experimental average (\ref{Eq:exp-av}) and the
predictions of chiral symmetry~\cite{ChPT}. When taking into account the
next-to-leading order in the chiral expansion, the prediction of chiral
perturbation theory for $\kappa_{\pi^\pm}$ is reduced from its one-loop value
$\kappa_{\pi^\pm}^{[1]}=(2.7\pm0.1)\times10^{-4}\;\mbox{fm}^3$ to the
two-loop result
$\kappa_{\pi^\pm}^{[2]}=(2.4\pm0.5)\times10^{-4}\;\mbox{fm}^3.$ The combined
(net) effect of these two shifts is that the value of $\kappa_{\pi^\pm}$
obtained by chiral perturbation theory is still by a factor 1.7 below the
experimental average. Considering the quoted errors, at present the
predictions of chiral symmetry (and therefore of low-energy QCD) cannot be
regarded as consistent with experiment.\item The results based on
quantum-field-theoretic descriptions of the pion
\cite{Bernard89,ChPT,Ivanov92,Koll01} lie within a factor 2.3 around the
geometrical central value $\kappa_{\pi^\pm}=5.5\times
10^{-4}\;\mbox{fm}^3$:$$\kappa_{\pi^\pm}^{\rm QFT}=5.5\times 2.3^{\pm1}\times
10^{-4}\;\mbox{fm}^3\ .$$Thus these approaches yield the correct order of
magnitude of the polarizability. \item In contrast to this, a nonrelativistic
(potential-model) description \cite{Leeb} of the~pion as ${\rm q}\bar{\rm q}$
bound state yields for the polarizability $\kappa_{\pi^\pm}$ of the charged
pion a numerical value which is~by a factor 80 smaller than the experimental
average (\ref{Eq:exp-av}).\end{itemize} Consequently, we find a rather large
discrepancy between the two classes of theoretical approaches to the pion
discussed here. In view of this spread of predictions, it does~not make much
sense to compute the average of all theoretical predictions listed in
Table~\ref{Tab:Theory}.

\subsection{Motivation}The brief confrontation of theory and experiment, in
Subsection~\ref{Subsect:TVE}, demonstrates that the theoretical prediction
for the electric polarizability of charged pions computed in~a
nonrelativistic quark model is smaller than the experimental average
(\ref{Eq:exp-av})~by almost~two orders of magnitude \cite{Leeb}. In view of
this, the authors of Ref.~\cite{Leeb} arrive at the conclusion that such
potential models can just grasp the effect of the ``hard core'' of
valence~quarks but miss completely the by far dominant contribution of the
``cloud'' of virtual~mesons.

In order to bridge the gulf between the theoretical approaches mentioned
above,~we study the electric polarizability of the (charged) pion in a {\em
semirelativistic\/} description~of hadrons. In particular, we would like to
find out whether the observed smallness of~the quark core contribution to the
electric polarizability of the pion is merely some artifact of the
nonrelativistic kinematics which is removed as soon as the relativistically
correct expression for the kinetic energy of the bound-state constituents is
taken into account.

Needless to say, we cannot expect that the increase of consistency brought
about~by a more relativistic treatment will be able to explain the {\em
total\/} polarizabilities of mesons: since we describe a hadron as a bound
state of constituent quarks, we can at most~hope to find a significant
increase of the quark core contribution to the meson polarizability.

Consequently, we do not even attempt to predict a theoretical value for the
absolute magnitude of the charged-pion electric polarizability. Rather, if
there is at~all a gain~for this polarizability by the inclusion of
relativistic corrections, we will quantify this~yield.

\section{(Static) Electric Polarizability of Charged Pions}Let us start the
present analysis from the assumption that the composite system under
consideration can be described by a Hamiltonian $H_0$ and all effects of
polarizability~can be summarized by some contribution $W$ to the total
Hamiltonian $H,$ which thus reads\begin{equation}H=H_0+W\
.\label{Eq:fullHam}\end{equation}

\subsection{Electric polarizability of mesons within quantum theory}As
mentioned already in the Introduction, an appropriate stage for the
investigation~of the electromagnetic polarizabilities of some particle $P$ is
the Compton scattering off~$P;$ the low-energy expansion of the Compton
scattering amplitude provides a definition~of the electromagnetic
polarizabilities of $P$ as the lowest nontrivial expansion coefficients.
Here, however, we follow the more intuitive (albeit static) picture drawn in
Refs.~\cite{Leeb,Lucha91}, which is based on the displacement of the quarks
constituting a meson in electric fields.

In this paper, we employ the Heaviside--Lorentz system of units for
electromagnetic quantities, i.e., for the electric charge $e$ as well as the
electric field strength $\mbox{\boldmath{$E$}}$ and the magnetic field
strength $\mbox{\boldmath{$B$}}$. In this system the electromagnetic fine
structure constant~$\alpha$ is related to the electric charge unit $e$
by$$\alpha\equiv\frac{e^2}{4\pi}\simeq\frac{1}{137.036}\ .$$(Note that, with
very few exceptions, all theoretical and experimental investigations~of
electromagnetic polarizabilities use Gaussian units, characterized by
$\alpha\equiv e^2\simeq 1/137.$ Appropriate factors $4\pi$ in those
electromagnetic relations which involve polarizabilities guarantee the
identity of the polarizabilities in Gaussian and Heaviside--Lorentz units.)

In order to extract the electric polarizability $\kappa$ of a composite
particle, one analyzes the response of this particle to an external electric
field. If the electric dipole moment $\mbox{\boldmath{$d$}}$ induced thereby
is proportional to the electric field strength $\mbox{\boldmath{$E$}}$ of
this external field, the electric polarizability $\kappa$ is defined as
constant of proportionality between $\mbox{\boldmath{$E$}}$
and~$\mbox{\boldmath{$d$}}$:$$\mbox{\boldmath{$d$}}
=4\pi\,\kappa\,\mbox{\boldmath{$E$}}\ .$$The shift $\Delta E$ of some energy
level $E$ resulting from this residual interaction is
given~by\begin{equation}\Delta
E=-\frac{1}{2}\,\mbox{\boldmath{$d$}}\cdot\mbox{\boldmath{$E$}}
=-2\pi\,\kappa\,\mbox{\boldmath{$E$}}^2\ .\label{Eq:ES-resint}\end{equation}
Thus the electric polarizability $\kappa$ may be read off from the
contribution of second~order in the electric field strength
$\mbox{\boldmath{$E$}}$ to the energy $E$ of the composite particle exposed
to~$\mbox{\boldmath{$E$}}.$

Assume that the meson under consideration is built up of a quark and an
antiquark, with masses $m_1$ and $m_2,$ and electric charges $Q_1$ and $Q_2,$
located at the coordinates~$\mbox{\boldmath{$r$}}_1$ and
$\mbox{\boldmath{$r$}}_2,$ respectively. Let the electric field
$\mbox{\boldmath{$E$}}$ be generated by a point-like electric charge of
magnitude $Z\,e,$ where $e$ denotes the magnitude of the electron charge. The
residual interaction due to the induced electric dipole moment may be found
by expanding~the potential energy $V_{\rm e}$ of the two quarks which
constitute the meson in this electric field,$$V_{\rm
e}=Z\,\alpha\left(\frac{Q_1}{|\mbox{\boldmath{$r$}}_1|}+
\frac{Q_2}{|\mbox{\boldmath{$r$}}_2|}\right),$$according to the relation
$$\frac{1}{|\mbox{\boldmath{$R$}}+\mbox{\boldmath{$r$}}|}\simeq\frac{1}
{|\mbox{\boldmath{$R$}}|}\left(1-
\frac{\mbox{\boldmath{$R$}}\cdot\mbox{\boldmath{$r$}}}
{|\mbox{\boldmath{$R$}}|^2}\right),$$over the center-of-momentum
coordinate$$\mbox{\boldmath{$R$}}\equiv
\frac{m_1\,\mbox{\boldmath{$r$}}_1+m_2\,\mbox{\boldmath{$r$}}_2}{m_1+m_2}$$of
the two-particle system forming the meson. Expressed in terms of the
corresponding relative coordinate$$\mbox{\boldmath{$r$}}\equiv
\mbox{\boldmath{$r$}}_1-\mbox{\boldmath{$r$}}_2$$of the two bound-state
constituents and the electric field strength $\mbox{\boldmath{$E$}}$
generated by the point-like electric charge $Z\,e,$
$$\mbox{\boldmath{$E$}}=\frac{Z\,e}{4\pi\,|\mbox{\boldmath{$R$}}|^3}\,
\mbox{\boldmath{$R$}}\ ,$$this potential energy $V_{\rm e}$ is thus given
by$$V_{\rm e}=Z\,(Q_1+Q_2)\,\frac{\alpha}{|\mbox{\boldmath{$R$}}|}+
\frac{m_1\,Q_2-m_2\,Q_1}{m_1+m_2}\,e\,
\mbox{\boldmath{$r$}}\cdot\mbox{\boldmath{$E$}}\ .$$It goes without saying
that one is only interested in the energy shift $\Delta E$ brought~about by
the possibility to polarize the hadron. Consequently, the perturbation $W$
due to~the polarization of this hadron induced by the Coulomb interaction of
the quarks with~the external point-like charge $Z\,e$ is given by
$$W=\frac{m_1\,Q_2-m_2\,Q_1}{m_1+m_2}\,e\,
\mbox{\boldmath{$r$}}\cdot\mbox{\boldmath{$E$}}\ .$$

We obtain the energy eigenvalues $E$ of the Hamiltonian $H$ by a standard
variational technique \cite{VT} around the normalized eigenstates
$|\phi_0\rangle$ of the unperturbed Hamiltonian $H_0,$ defined by
$$H_0\,|\phi_0\rangle=E_0\,|\phi_0\rangle\
,\quad\langle\phi_0|\phi_0\rangle=1\ ,$$where $E_0$ labels the bound-state
energies in the absence of the external electric
field~$\mbox{\boldmath{$E$}}.$ Introducing a real variational parameter
$\lambda=\lambda^\ast,$ we define a set $\{|\phi_\lambda\rangle\}$ of trial
states~by$$|\phi_\lambda\rangle=(1+\lambda\,W)\,|\phi_0\rangle\ .$$ Just for
notational simplicity, in the following all expectation values without
explicitly denoted states, that is, all expectation values written in the
form $\langle{\cal O}\rangle$ for some operator ${\cal O},$ have to be
understood to be evaluated with respect to the unperturbed
states~$|\phi_0\rangle$:$$\langle{\cal O}\rangle\equiv\langle\phi_0|{\cal
O}|\phi_0\rangle\ .$$

We only investigate the ground state of the composite system under
consideration; this state is characterized by vanishing orbital angular
momentum and thus spherically symmetric. We note that expectation values
$\langle W^{2n+1}\rangle,$ $n=0,1,\dots,$ of odd powers~of~$W$ with respect
to states corresponding to a vanishing orbital angular momentum
vanish:$$\langle W^{2n+1}\rangle=0\ ,\quad n=0,1,2,\dots\ .$$Hence, we find,
for the square of the norm of the states $|\phi_\lambda\rangle,$
$$\langle\phi_\lambda|\phi_\lambda\rangle=1+\lambda^2\,\langle W^2\rangle$$
and, for $\langle\phi_\lambda|H|\phi_\lambda\rangle,$
$$\langle\phi_\lambda|H|\phi_\lambda\rangle=E_0+2\,\lambda\,\langle
W^2\rangle+\lambda^2\,\langle W\,H_0\,W\rangle\ .$$The expectation values of
the Hamiltonian $H$ with respect to the above trial
states~$|\phi_\lambda\rangle$ yields a set of variational approximations
$E(\lambda)$ to the exact energy eigenvalue $E$ of~$H$:
\begin{eqnarray*}
E(\lambda)&\equiv&\frac{\langle\phi_\lambda|H|\phi_\lambda\rangle}
{\langle\phi_\lambda|\phi_\lambda\rangle}\\[1ex] &=&E_0+2\,\lambda\,\langle
W^2\rangle+\lambda^2\,(\langle W\,H_0\,W\rangle-E_0\,\langle
W^2\rangle)+{\cal O}(W^4)\\[1ex] &=&E_0+2\,\lambda\,\langle
W^2\rangle+\lambda^2\,\langle [W,H_0]\,W\rangle+{\cal O}(W^4)\
.\end{eqnarray*}The minimum energy eigenvalue $\bar E$ is determined by the
requirement of stationarity~of the latter expression with respect to the
variational parameter $\lambda$:$$\bar E=E_0-\frac{\langle
W^2\rangle^2}{\langle[W,H_0]\,W\rangle}\ .$$The second term on the right-hand
side of this result represents the energy shift $\Delta E.$

Moreover, for spherically symmetric states, the expectation value in the
numerator of the above energy shift may be simplified with the help of
$$\langle(\mbox{\boldmath{$r$}}\cdot\mbox{\boldmath{$E$}})^2\rangle
=\frac{1}{3}\,\mbox{\boldmath{$E$}}^2\,\langle r^2\rangle\ ,\quad
r\equiv|\mbox{\boldmath{$r$}}|\ .$$A final simplification of this energy
shift is achieved by choosing our spatial coordinate frame such that the
(external) electric field strength $\mbox{\boldmath{$E$}}$ is aligned along
the $z$-axis of~our coordinate frame;~this choice entails
$$\mbox{\boldmath{$r$}}\cdot\mbox{\boldmath{$E$}}=z\,|\mbox{\boldmath{$E$}}|\
.$$When comparing the energy shift thus obtained within the above variational
technique with the ansatz (\ref{Eq:ES-resint}), we find that the electric
polarizability $\kappa$ of mesons is given by~the general expression
\begin{equation}\kappa=\frac{2}{9}\left(\frac{m_1\,Q_2-m_2\,Q_1}{m_1+m_2}
\right)^2\alpha\,\frac{\langle r^2\rangle^2}{\langle[z,H_0]\,z\rangle}\
.\label{Eq:ElPol}\end{equation}

\subsection{Relativistic kinematics: the ``spinless Salpeter
equation''}\label{Subsect:RelKin-SSE}Let us perform for the unperturbed
Hamiltonian $H_0$ introduced in Eq.~(\ref{Eq:fullHam}) the first step in an
attempt to reconcile the potential-model approach to bound states of
quarks~with relativity. The semirelativistic Hamiltonian $H_0$ governing the
dynamics of two particles with masses $m_1$ and $m_2$ and relative momentum
$\mbox{\boldmath{$p$}}$ in their center-of-momentum frame involves the
relativistic kinetic energies $T_i(p),$ $i=1,2,$ of the bound-state
constituents,$$T_i(p)\equiv\sqrt{\mbox{\boldmath{$p$}}^2+m_i^2}\ ,\quad
p\equiv|\mbox{\boldmath{$p$}}|\ ,\quad i=1,2\ ,$$and the relevant interaction
potential $V(\mbox{\boldmath{$r$}}),$ which depends on the relative
coordinate~$\mbox{\boldmath{$r$}}$ of the bound-state constituents and is
responsible for the formation of the bound~state:
\begin{equation}H_0=T_1(p)+T_2(p)+V(\mbox{\boldmath{$r$}})\
.\label{Eq:SHam2}\end{equation}In the case of mesons, this potential
$V(\mbox{\boldmath{$r$}})$ arises from the strong interactions described,
according to our common understanding, by quantum chromodynamics:
$V(\mbox{\boldmath{$r$}})=V_{\rm s}(r).$

The eigenvalue equation for a Hamiltonian $H$ which is the sum of the
``square-root'' operator of the relativistic expression for the free (i.e.,
kinetic) energies of the~particles constituting the physical system under
consideration and a static interaction potential is called the ``spinless
Salpeter equation.'' It may be obtained from the Bethe--Salpeter equation
\cite{Salpeter51}---which describes bound states in (relativistic) quantum
field theory---by assuming the interaction to be instantaneous (which yields
the Salpeter equation~\cite{Salpeter52}) and by neglecting all spin degrees
of freedom of the involved bound-state constituents. On the other hand, it
can be regarded as the simplest generalization of the Schr\"odinger equation
of standard nonrelativistic quantum theory towards relativistic kinematics.

By substituting the unperturbed Hamiltonian $H_0$ of Eq.~(\ref{Eq:SHam2})
into our general result (\ref{Eq:ElPol}) (and by taking into account that
$[z,V(\mbox{\boldmath{$r$}})]=0$), we find that here the magnitude of the
electric polarizability $\kappa$ is determined by the following ratio of
expectation values:$$R\equiv\frac{\langle r^2\rangle^2}
{\langle[z,H_0]\,z\rangle}=\frac{\langle r^2\rangle^2}
{\langle[z,T_1(p)+T_2(p)]\,z\rangle}\ .$$The question is whether this ratio
$R$ is greater or less than its nonrelativistic limit~$R_{\rm NR}.$

Let the ground state of the meson under consideration be described by a real
wave function $\psi(p).$ The evaluation of the expectation value
$\langle[z,T_i(p)]\,z\rangle$ is straightforward:\begin{eqnarray*}
\langle[z,T_i(p)]\,z\rangle&=&\frac{2\pi}{3}\int\limits_0^\infty{\rm
d}p\,p\left(2\,\frac{\partial T_i(p)}{\partial
p}+p\,\frac{\partial^2T_i(p)}{\partial p^2}\right)\psi^2(p)\\[1ex]
&=&\frac{1}{6}\int{\rm
d}^3p\,\frac{2\,p^2+3\,m_i^2}{\left(p^2+m_i^2\right)^{3/2}}\,\psi^2(p)\
.\end{eqnarray*}It is rather easy to convince oneself that the expectation
value $\langle[z,T_i(p)]\,z\rangle$ is bounded from above by its
nonrelativistic limit $\langle[z,T_i(p)]\,z\rangle_{\rm NR}$:
$$\langle[z,T_i(p)]\,z\rangle\le\langle[z,T_i(p)]\,z\rangle_{\rm
NR}=\frac{1}{2\,m_i}\ .$$This expectation value enters into the denominator
of $\kappa.$ Hence, this inequality tends to increase the ratio $R$ and the
value of $\kappa$ compared with the nonrelativistic situation. Note that---in
contrast to the general semirelativistic case---in the nonrelativistic limit
the expectation value $\langle[z,T_i(p)]\,z\rangle$ reduces to a constant
(namely, the inverse of $2\,m_i$), that is, it depends no longer on the wave
function $\psi$ which describes the bound state.

Within the present approach it suffices to model the hadrons by a static,
spherically symmetric interaction potential $V_{\rm s}(r).$ For recent
reviews of the description of hadrons as bound states of quarks in terms of
both nonrelativistic and semirelativistic potential models see, for instance,
Refs.~\cite{Lucha91,Lucha92}. The prototype of a large class of
quark--antiquark interaction potentials is the ``Coulomb-plus-linear'' (or,
in view of its shape, ``funnel'') potential. This static potential is a
linear combination of a Coulomb term (arising~from one-gluon exchange between
the strongly interacting quarks) and a linearly rising term (which is assumed
to summarize all nonperturbative effects of the strong interactions):
\begin{equation}V_{\rm s}(r)=-\frac{4}{3}\,\frac{\alpha_{\rm s}}{r}+a\,r\
.\label{Eq:FP}\end{equation}

For the actual application of the semirelativistic quark model introduced
above, we have to specify the numerical values of the parameters entering in
the Hamiltonian~$H_0.$ In the case of the pion, we obviously deal with
bound-state constituents of equal~masses $m=m_1=m_2.$ For this common mass,
we adopt the canonical value for the constituent mass of light nonstrange
quarks: $m=m_{\rm u}=m_{\rm d}=0.336\;\mbox{GeV}$ \cite{Lucha91}. The two
parameters of the interaction potential $V_{\rm s}(r)$ are taken from a
(nonrelativistic) fit \cite{Ono} of the meson spectrum. We use, for the
strong fine structure constant $\alpha_{\rm s}$ characterizing the coupling
strength of the Coulomb-like term in the funnel potential, the value
$\alpha_{\rm s}=0.31$ and,~for the slope $a$ of the linear contribution to
the funnel potential, the value $a=0.15\;\mbox{GeV}^2.$ (One might be tempted
to question the choice of a rather~large constituent quark~mass $m$ for the
description of the comparatively light pion. However,~one has to bear in~mind
that the numerical values of the mass and coupling parameters used here
have~emerged from a satisfactory simultaneous fit \cite{Ono} of both the
meson and the baryon mass~spectra within a specific {\em nonrelativistic\/}
quark-potential model. It goes without saying that~any semirelativistic
description of hadrons will require slightly different numerical~values~of
the parameters involved. In principle, all parameters, that is, the quark
mass $m$ and the two couplings $\alpha_{\rm s}$ and $a,$ should be readjusted
for the present {\em semirelativistic\/} treatments of bound states of
quarks. Moreover, the interquark potential employed in Ref.~\cite{Ono} is~of
a more complicated parametric shape than the simple funnel potential $V_{\rm
s}(r)$ in~Eq.~(\ref{Eq:FP}). It appears, however, very unlikely that
such---comparatively minor---modifications are able to induce drastic changes
in the resulting predictions for the electric polarizability. At least, they
will hardly alter the order of magnitude of the computed polarizability.)

The discrete eigenvalues and corresponding eigenstates of the operator
represented by the semirelativistic Hamiltonian $H_0$ of Eq.~(\ref{Eq:SHam2})
are (approximately) determined with the help of the well-known
minimum--maximum principle~\cite{MMP}. The trial space required by this
technique is spanned by the ``Laguerre'' basis states defined in
Refs.~\cite{Lucha97,Lucha98O,Lucha98D}.

The relativistic virial theorem derived in Ref.~\cite{Lucha89:RVT} (for a
very comprehensive review, see Ref.~\cite{Lucha90:RVTs}) allows to define a
precise quantitative measure \cite{Lucha99Q,Lucha99A} for the quality~of the
results obtained within the framework of variational techniques. According to
the analysis presented in Refs.~\cite{Lucha99Q,Lucha99A}, for a generic
Hamiltonian operator $H$ consisting~of a (momentum-dependent) kinetic term
$T(\mbox{\boldmath{$p$}})$ and a (coordinate-dependent) interaction potential
$V(\mbox{\boldmath{$x$}}),$ that
is,$$H=T(\mbox{\boldmath{$p$}})+V(\mbox{\boldmath{$x$}})\ ,$$the accuracy of
a trial state $|\varphi\rangle$ which approximates the exact bound state
under~study may be quantitatively estimated by the deviation from zero of the
quantity
$$\nu\equiv\frac{\langle\varphi|\mbox{\boldmath{$p$}}\cdot\frac{\partial}
{\partial\mbox{\boldmath{$p$}}}T(\mbox{\boldmath{$p$}})|\varphi\rangle}
{\langle\varphi|\mbox{\boldmath{$x$}}\cdot\frac{\partial}
{\partial\mbox{\boldmath{$x$}}}V(\mbox{\boldmath{$x$}})|\varphi\rangle}-1\
.$$The main advantage of this measure for the accuracy of approximate
eigenstates $|\varphi\rangle$ is that it does not require any information on
the solutions of the investigated eigenvalue problem other than the one
provided by the variational approximation technique itself.

Clearly, any enlargement of the dimension $d$ of the minimum--maximum trial
space will, in general, improve the obtained approximation. For our choice of
$d=25$ (that~is, when truncating the expansion of the approximate eigenstates
$|\varphi\rangle$ after 25 basis~states) we find for the above quality
measure $\nu$ the numerical value $\nu=2\cdot10^{-4}.$ Beyond~doubt, the
precision achieved in this way may be regarded as sufficient for the present
purpose.

In order to estimate the effects of relativistic kinematics, we compare the
outcome of the above semirelativistic treatment with the corresponding
nonrelativistic~solution. The nonrelativistic counterparts of the
semirelativistic eigenvalues and eigenstates are found with rather high
accuracy by solving the Schr\"odinger equation with a (standard) numerical
integration procedure developed in Ref.~\cite{Lucha99M} precisely for this
purpose.

\begin{table}[h]\caption{Expectation values $\langle r^2\rangle$ and
$\langle[z,H_0]\,z\rangle$ determining the electric polarizabilities of
mesons, obtained within semirelativistic and nonrelativistic descriptions of
the pion, for a funnel-shaped quark interaction potential $V_{\rm
s}(r)=-\frac{4}{3}\,\alpha_{\rm s}/r+a\,r$ with a light-quark constituent
mass $m=0.336\,\mbox{GeV}$ and coupling constants $\alpha_{\rm s}=0.31$ and
$a=0.15\,\mbox{GeV}^2.$}\label{Tab:PiPol}
\begin{center}\begin{tabular}{ccc}\hline\hline\\[-1.5ex]
\multicolumn{1}{c}{Kinematics}&\multicolumn{1}{c}{$\langle r^2\rangle$}&
\multicolumn{1}{c}{$\langle[z,H_0]\,z\rangle$}\\[1ex]
\multicolumn{1}{c}{}&\multicolumn{1}{c}{$[\mbox{GeV}^{-2}]$}&
\multicolumn{1}{c}{$[\mbox{GeV}^{-1}]$}\\[1ex]\hline\\[-1.5ex]
relativistic&11.7&1.66\\[1ex]nonrelativistic&17.8&2.98
\\[1ex]\hline\hline\end{tabular}\end{center}\end{table}

The resulting expectation values $\langle r^2\rangle$ and
$\langle[z,H_0]\,z\rangle,$ obtained for both relativistic and
nonrelativistic kinematics along the lines sketched above, are presented in
Table~\ref{Tab:PiPol}. These two expectation values determine the ratio $R$
and, in consequence of the general result~(\ref{Eq:ElPol}), the theoretical
prediction for the (static) electric polarizability $\kappa$ of mesons.
Insertion of the numerical values of Table~\ref{Tab:PiPol}, however, leads to
the (disappointing) ratio$$\frac{\kappa}{\kappa_{\rm NR}}=\frac{R}{R_{\rm
NR}}=\frac{\langle r^2\rangle^2}{\langle r^2\rangle^2_{\rm NR}}\,
\frac{\langle[z,H_0]\,z\rangle_{\rm NR}}{\langle[z,H_0]\,z\rangle}=0.77<1\
.$$This tells us that merely replacing, in the Hamiltonian $H,$
nonrelativistic by relativistic kinematics reduces the valence-quark
contribution to the pion's electric polarizability.

\subsection{Relativistic corrections to the static interaction potential}The
static potential may be regarded as the lowest term in a (relativistic)
expansion of the interaction of the bound-state constituents over the
nonrelativistic limit. Of course, one might ask whether the inclusion of (the
totality of all) relativistic corrections to~the interaction potential will
save the day. The derivation of the relativistic corrections to the static
interaction potential from the corresponding two-particle scattering problem
has been thoroughly analyzed in
Refs.~\cite{Lucha91RelTreat,Lucha92Comment,Lucha91Regge} and extensively
reviewed in Ref.~\cite{Lucha92}.

Since the pion is a quark-antiquark bound state corresponding to vanishing
relative orbital angular momentum of its constituents, only the spin-spin
term will contribute. Moreover, in the ultrarelativistic case of {\em
massless\/} bound-state constituents, that is,~for $m_1=m_2=0,$ we encounter
a comparatively simple situation. In this case, at least, it is possible to
make analytic statements about the influence of semirelativistic treatments
of both kinetic terms and interaction-potential terms on the pion electric
polarizability. To this end, we take advantage of explicit results presented
in Section~VII of Ref.~\cite{Lucha91RelTreat}. Within the variational
evaluation employed for the example studied there, it is easy~to convince
oneself that, for the expectation value of $r^2$ entering into the numerator
of~the ratio $R,$ the value $\langle r^2\rangle_{\rm RC}$ obtained from
semirelativistic Hamiltonians that include the relativistic corrections to
the interaction potential is always smaller than the result~$\langle
r^2\rangle$ derived from the corresponding operators with just the static
potential: $\langle r^2\rangle_{\rm RC}\le\langle r^2\rangle.$ A closer
inspection reveals that, expressed in terms of the coupling parameters
$\alpha_{\rm s}$ and~$a$ of the funnel potential~(\ref{Eq:FP}), these
quantities differ, for Hydrogen-like trial functions,~by$$\langle
r^2\rangle-\langle r^2\rangle_{\rm RC}=\frac{64\,\alpha_{\rm
s}}{3\,\pi^2\,a}>0\ ;$$this may be translated into the inequality$$\langle
r^2\rangle_{\rm RC}\le\left(1-\frac{2\,\alpha_{\rm s}}{\pi}\right)\langle
r^2\rangle\ .$$Concerning the expectation value $\langle[z,H_0]\,z\rangle$
entering in the denominator of the ratio~$R$ as well as the more general case
of nonvanishing values of the bound-state constituents' masses, we did not
succeed to derive rigorous analytical statements on the influence~of
relativistic corrections to the interaction potential on the pion's electric
polarizability. A straightforward, entirely numerical, phenomenological
analysis shows that inclusion of these relativistic corrections entails a
significant further reduction of the value of the pion electric
polarizability compared with the spinless-Salpeter value of
Subsection~\ref{Subsect:RelKin-SSE}.

\section{Conclusions}This analysis has been devoted to the determination of
the valence-quark contribution to the electric polarizability of mesons (for
the experimentally most relevant particular example of the charged pion
$\pi^\pm$) within the framework of semirelativistic quark models based on a
conventional Hamiltonian description of hadrons as bound states of quarks.
Recently, one observes a revival of interest in the experimental study of the
electric~and magnetic polarizabilities of mesons via processes such as, e.g.,
the Primakoff effect,~cf.\ the proposal for the NA58 (COMPASS) experiment at
the CERN SPS collider \cite{COMPASS-P, COMPASS-A}. This motivates the
continued search for a satisfactory theoretical understanding of~this rather
fundamental quantity and the present attempt to reconcile different
approaches.

From this investigation we are forced to conclude that even within a
semirelativistic description of hadrons the valence-quark contribution
constitutes only a minor fraction of the total electric polarizability of
mesons. This point of view is, in fact, supported~by a study~\cite{Weiner85}
of the {\em nucleon\/} electromagnetic polarizabilities within a chiral quark
model. This model considers any hadron explicitly as a core formed by the
appropriate valence quarks which is surrounded by clouds of virtual hadrons,
in particular, of virtual~pions.

\end{document}